\documentclass[12pt,preprint]{aastex}

\shorttitle{LBGs at $z\sim5$: Rest-Frame UV Spectra}
\shortauthors{Ando et al.}

\begin{document}

\title{Lyman Break Galaxies at $z\sim5$: Rest-Frame UV Spectra\altaffilmark{1}}

\author{Masataka Ando\altaffilmark{2}, Kouji Ohta\altaffilmark{2}, Ikuru Iwata,\altaffilmark{2}}
\author{Chisato Watanabe\altaffilmark{2}, Naoyuki Tamura,\altaffilmark{3}}
\author{Masayuki Akiyama\altaffilmark{4}, and Kentaro Aoki\altaffilmark{4}}

\altaffiltext{1}{Based on data collected at Subaru Telescope, which is operated by the National Astronomical Observatory of Japan.}
\altaffiltext{2}{Department of Astronomy, Kyoto University, Kyoto 606-8502, Japan.}
\altaffiltext{3}{Department of Physics, University of Durham, Durham DH1 3LE, UK.}
\altaffiltext{4}{Subaru Telescope, National Astronomical Observatory of Japan, 650 North A'ohoku Place, Hilo, Hawaii 96720 USA.}

\begin{abstract}
We report initial results for spectroscopic observations of
candidates of Lyman Break Galaxies (LBGs) at $z\sim5$ in a region
centered on the Hubble Deep Field-North by using the Faint Object Camera
and Spectrograph attached to the Subaru Telescope.
Eight objects with $I_C\leq25.0$ mag, including one AGN, 
are confirmed to be at $4.5<z<5.2$. 
The rest-frame UV spectra of seven LBGs commonly show no or weak
Ly$\alpha$ emission
line (rest-frame equivalent width of $0-10$\AA) and relatively strong
low-ionization interstellar metal absorption lines of \ion{Si}{2}
$\lambda$1260, \ion{O}{1}+\ion{Si}{2}
$\lambda$1303, and \ion{C}{2} $\lambda$1334 
(mean rest-frame equivalent
 widths of them are $-1.2 \sim -5.1 $\AA).
These properties are significantly different from those of the mean
rest-frame UV spectrum of LBGs at $z\sim3$, 
but are quite similar to those of subgroups of LBGs at
$z\sim3$ with no or weak Ly$\alpha$ emission.
The weakness of Ly$\alpha$ emission and strong low-ionization
interstellar metal absorption lines may indicate that these LBGs at $z\sim5$
are chemically evolved to some degree and have a dusty environment.
Since the fraction of such LBGs at $z\sim5$ in our sample is larger than
that at $z\sim3$, we may witness some sign of evolution of LBGs 
from $z\sim5$ to $z\sim3$, though the present sample size is very small.
It is also possible, however, that the brighter LBGs tend to show no
or weak Ly$\alpha$ emission,
because our spectroscopic sample is bright (brighter than $L^{\ast}$)
among LBGs at $z\sim5$.
More observations
are required to establish spectroscopic nature of LBGs at $z\sim5$.
\end{abstract}

\keywords{cosmology: observations --- galaxies: active --- galaxies:
evolution --- galaxies: formation --- galaxies: high-redshift}

\section{Introduction}
In order to understand formation and evolution of galaxies,
systematic searches for galaxies in the early universe and detailed
studies of the detected galaxies are necessary.
In this decade, a number of galaxies detected at high redshift
($z>2\sim3$) has been rapidly increasing.
The use of photometric redshift selection from deep broad-band
images, especially so-called "Lyman break method" (e.g., \citet{Ste92}; Steidel, Pettini \& Hamilton 1995) gives the largest
sample of galaxies at $z\sim3$ \citep[Lyman Break Galaxies; LBGs,
e.g.,][]{Ste03}. 
Extensive studies  of them have been revealing 
individual and statistical nature of star-forming galaxies at $z\sim3$
(e.g., \citet{Ste99,Ste98,Pet01}; Papovich, Dickinson, \& Ferguson 2001).

At the same time, their optical follow-up spectroscopy
has also been made extensively \citep[e.g.,][]{Ste96a,Ste96b,Ste99},
and revealed rest-frame UV spectral features of LBGs in addition to their
redshift information. 
Using about 800 spectra of LBGs at $z\sim3$, \citet{Shap03} found 
that about three fourth of them show significant Ly$\alpha$ emission line and the 
remainders show only Ly$\alpha$ absorption.
The LBGs with weaker Ly$\alpha$ emission
have stronger low-ionization interstellar metal absorption lines and 
redder UV continua.
These trend suggest that the LBGs at $z\sim3$ with weak Ly$\alpha$ 
emission are more metal enriched and dusty, i.e., chemically evolved
than those with strong Ly$\alpha$ emission.

How about properties of LBGs at higher redshift?
Are there any signs of evolution of LBGs compared with LBGs at $z\sim3$? 
To answer the question and obtain clues to understand formation and
evolution of galaxies in the early universe, we made a
systematic search for LBGs at $z\sim5$ ($\sim$1Gyr earlier to $z=3$).
We carried out wide (effectively $\sim600$ arcmin$^2$) and deep broad-band ($V,
I_C,$ and $z'$) imaging
observations toward an area centered on the Hubble Deep Field-North
\citep[HDF-N;][]{Wil96} with Suprime-Cam \citep{Miya} attached to
the Subaru telescope.
Thanks to plenty of redshift information of galaxies in and around 
the HDF-N, we could set suitable color criteria on the 
two-color ($V-I_C$ and $I_C-z'$) diagram to effectively select  galaxies 
at $4.5\lesssim z \lesssim5.5$ avoiding foreground contamination, and
obtained 
$\sim300$ LBG candidates at $z\sim5$ with $23.5$ mag
$<I_C\leq26.0$ mag \citep{Iwa03}. 
This is one of the largest LBG samples at $z\sim5$
systematically surveyed.
Using this sample, we statistically derived a rest-frame UV luminosity
function (UVLF) of LBGs at $z\sim5$, which
shows only a slight change from $z\sim5$ to $z\sim3$.
We also found that a spatial distribution of LBGs at $z\sim5$ 
shows a large void-like structure and
a larger clustering amplitude than that at $z\sim3$
(in preparation).
The presence of a similar void-like structure is seen in 
\citet{Bre04}, and a large clustering amplitude of LBGs at $z\sim5$ may also 
suggest the presence of the larger bias in galaxy formation at $z\sim5$
than at $z\sim3$ \citep{Ouchi03c}. 

In order to confirm redshifts of the LBG candidates and to study
spectral features and relations with other properties of them,
we started follow-up optical spectroscopy of our LBG candidates at $z\sim5$.
Although similar systematic searches for $z=5\sim6$ galaxies
have been carried out 
(e.g., \citet{Ouchi03b,Bou03}; Yan, Windhorst, \& Cohen 2003;
Stanway, Bunker, \& McMahon 2003),
a number of spectra of LBGs at $z\sim5$ is  still extremely small.
Moreover most of the spectra obtained so far show only Ly$\alpha$
emission,
and the continuum spectral feature and relations with
other properties observed in LBGs at $z\sim3$ are
still quite unknown.
In this paper, we present initial results of the optical
spectroscopy of a part of our sample.
Observations and data reduction are described in section 2, and the
results are presented in section 3. 
In section 4, we compare the results  with spectral features obtained
for LBGs at $z\sim3$, and discuss possible origins of the
similarity and difference seen in spectral features. 
The magnitude system is based on \rm{AB} magnitude unless otherwise
noted.

\section{Observations and Data Reduction}

We made optical spectroscopy for a subset of our sample using
multi-object-spectroscopy (MOS) mode of the Faint Object Camera and
Spectrograph \citep[FOCAS;][]{kashi} attached to the 8.2-m Subaru Telescope. 
Main spectroscopic targets are our LBG candidates brighter than $I_C =
25.0$ mag.
Since a mask plate of FOCAS MOS covers a 6$^{\prime}$ aperture diameter
field of view, 
we designed MOS masks to contain as many main targets as possible
on each MOS field. 
The number of targets thus selected is 17. 
In addition to them,
we selected 7 objects with $I_C < 25.0$ mag which lie out of the color
selection window but near the border line of the window
in order to examine our color selection criteria.
We also included fainter LBG candidates ($I_C \geq 25.0$ mag)
as many as possible in each mask.

Spectroscopic observations were made on 2003 February 24-26 under clear
sky conditions.
We used the grism of 300 lines/mm blazed at 7500\AA\ and
the SO58 order cut filter. 
The wavelength coverage was from 5800\AA\ to 10000\AA\ 
(depending on a slit position on a mask) with a pixel scale of 1.34\AA. 
The slit lengths were typically 10$^{''}$, and the slit widths were 
fixed to be 0.$^{''}$8, giving a spectral resolution of R$\sim$500.
One CCD pixel covered 0.$^{\prime\prime}$1 and 
a spatial sampling
was 0.$^{''}$3 pixel$^{-1}$ by on-chip three pixel binning.
An exposure time of each frame was 0.5 hours, and a total exposure
time was 5.5 hours for two masks (mask name F01 and F02), 
and 5.0 hours for one mask (mask name F08).
We nodded the telescope with $\sim$1$^{\prime\prime}$ along the slit
length for each exposure.
Spectrophotometric standard stars Feige 34 and Feige 66 were observed
with long-slit mode (a slit width of 2$^{\prime\prime}$) for 
sensitivity correction. 
Seeing during the observing runs was 0.$^{\prime\prime}5 -
0.^{\prime\prime}$8. 

The data were reduced with standard procedure using IRAF\footnote{Image
Reduction and Analysis Facility, distributed by National Optical
Astronomical Observatories, which are operated by the Association of
Universities for Research in Astronomy, Inc., under cooperative
agreement with the National Science Foundation.}. 
Bias was subtracted by
using the overscan region and bias frames, and flat-fielding was made by
normalizing averaged dome flat images. 
The spectra for each target were
then carefully aligned and combined for each night. 
Wavelength calibration was made using night sky emission lines exposed in the
object frame with rms errors of $0.4-0.9$ \AA. 
Sky emission was subtracted by using BACKGROUND task of IRAF. 
Five pixels were binned for a wavelength direction to improve S/N. 
One-dimensional spectrum of each object was extracted by
using APALL task of IRAF. 
Since our targets are faint, an aperture for the extraction was determined
for each object by eyes
to trace the object well. 
After sensitivity correction (including correction for atmospheric
A and B bands  by tracing the absorption features 
in the spectra of the standard stars) was applied to combined spectra 
for each night, we obtained final spectra by combining them.

\section{Results}
Figure 1 shows spectra of eight objects identified to be at $z\sim5$.
Object No.3 (F01-03) has already been identified with an AGN at $z=5.186$ 
by \citet{Bar02} in a follow-up spectroscopy of 1Ms {\it Chandra}
observation at the HDF-N. 
A significant continuum break at the redshifted Ly$\alpha$ is seen
in all the spectra.
The average of the depression factor $D_A$ \citep{Oke82} for 
these objects is $\sim0.6$ which is very close to those
for QSOs at $z\sim5$ \citep[e.g.,][]{Son02} and the estimation by
\citet{Mad95}. 
The spectra of seven LBGs (excluding object No.3)
also show low ionization interstellar (LIS) metal absorption lines such as 
\ion{Si}{2} $\lambda$1260, \ion{O}{1} +\ion{Si}{2} $\lambda$1303, and
\ion{C}{2} $\lambda$1334  
which are prominent features in UV spectra of nearby starburst
galaxies \citep[e.g.,][]{Hek98} and  in  spectra of LBGs
at $z\sim3$ \citep[e.g.,][]{Ste96a,Shap03}.
Thus they are securely identified to be at $z\sim5$.
We determined their redshifts as an average of these two or three
LIS absorption lines and listed the results in Table 1
as well as their coordinates and photometric data.
The obtained redshift range from 4.5 to 5.2
is consistent with that expected from the color selection,
Figure 7 of \citet{Iwa03}.
We also constructed a composite rest-frame spectrum of the seven LBGs
by scaling the continuum level of each spectrum and 
show the result in Figure 2, together with the composite spectrum of 811
LBGs at $z\sim3$ \citep{Shap03}.

Intriguingly, spectra of the seven LBGs in Figure 1 show no or weak
Ly$\alpha$ emission\footnote{
Object No.3 (F01-03) is an AGN and shows a broad ($\sim1000$ km$^{-1}$)
 and strong (EW$_{\rm rest}\sim44$\AA) Ly$\alpha$ emission.}.
Four objects show no significant Ly$\alpha$ emission line, and
the other three objects, No.1, No.6, and No.7 show a Ly$\alpha$ emission line,
but their rest-frame equivalent widths are rather weak;
EW$_{\rm rest}\sim1$\AA, $\sim6$\AA, and $\sim10$\AA, respectively,
with an error of 20--30\%.
The average rest-frame equivalent width of Ly$\alpha$ of the seven LBGs
is 2.5 \AA, and that derived from the composite spectrum is 4.5\AA
\footnote{The S/N of the continuum is improved 
much in the composite spectrum.
This leads to lower the continuum level in a longer wavelength region
adjacent to the Ly$\alpha$ line, which seem to result in the slightly larger
Ly$\alpha$ equivalent width for the composite spectrum.}.
The position of Ly$\alpha$ for some of them locates in the 
wavelength region of the atmospheric B band absorption.
We estimate the uncertainty in the correction for the absorption is
at most $\sim10$\%.

A mean rest-frame equivalent width of three LIS
absorption lines (\ion{Si}{2} $\lambda$1260, \ion{O}{1}+\ion{Si}{2}
$\lambda$1303, and \ion{C}{2} $\lambda$1334) for each of the seven LBGs is
measured to be $-1.2 \sim -5.1$\AA\ and is listed in Table 1.
The average rest-frame equivalent widths of \ion{Si}{2} $\lambda$1260,
\ion{O}{1}+\ion{Si}{2} 
$\lambda$1303, and \ion{C}{2} $\lambda$1334 of the seven LBGs are 
$-3.2$\AA, $-2.3$\AA, and $-2.4$\AA, respectively,
which agree with those measured in the composite spectrum;
$-2.8$\AA, $-2.0$\AA, and $-2.4$\AA, respectively\footnote{For some
objects, the LIS line falls in the wavelength region of the atmospheric A band,
and uncertainty of the correction is estimated to be $\sim 20$\%.}.
Some of the LBGs show a high ionization absorption feature of
\ion{Si}{4} $\lambda \lambda$1393, 1402  which are also detected
in local starburst galaxies and LBGs at $z\sim3$.
Most of them lie in the wavelength region where
the sky emission is strong or an atmospheric absorption is severe,
and we did not measure an equivalent width for each object.
The \ion{Si}{4} $\lambda\lambda$1393, 1402 feature can be seen in the
composite spectrum, thanks to the improvement of the S/N.

Peaks of the Ly$\alpha$ emission lines are redshifted with
respect to  the LIS lines 
by $650 \pm 300$ km s$^{-1}$, $530 \pm 300$ km s$^{-1}$, and
$700 \pm 300$ km s$^{-1}$, 
for No.1, No.6, and No.7, respectively.
In the composite spectrum, the offset is
$620 \pm 250$ km s$^{-1}$.
The value is almost the same as that of
the LBGs at $z\sim3$ with weak or no Ly$\alpha$ emission lines 
(630 km s$^{-1}$ for Group 2 by \citet{Shap03}) 
and also fits a relation between the outflow velocity and the equivalent
widths of LIS for LBGs at $z=4-5$ by Frye, Broadhurst, \& Benitez (2002).

We also measured a continuum slope $\beta$
($f_{\lambda}  \propto \lambda^{\beta}$) of the composite spectrum
to be $\beta \sim -0.55$ with a fitting uncertainty of $\pm0.3$.
However, since the value is determined in a short wavelength coverage
($\sim$250\AA) and depends on an applied binning scale and a clipping
threshold in making the composite spectrum, the uncertainty of $\beta$
would be much larger. 
Thus we do not discuss the result further.

Among the remainders of the spectroscopic sample in the color selection
window with $I_C < 25.0$ mag,
two objects show hints of the presence of
a continuum break at the wavelength corresponding to a redshifted
Ly$\alpha$ at $z\sim5$ 
and of one or two LIS absorption lines which
could be identified to be at $z\sim5$.
However, the S/Ns of their spectra are poor 
and we do not discuss these objects in this paper.
For the remaining targets, we could not find any significant features
due to the poor S/N.
A roughly estimated upper limit on the rest-frame Ly$\alpha$ equivalent width
for these objects is $\sim 20$ \AA. 
For the targets with $I_C \geq 25$ mag, we could not find any significant 
features due to the poor S/N.
The S/N of the spectrum depends not only on the $I_C$ magnitude,
but also on its surface brightness.
The objects identified to be at $z\sim5$ tend to have the sizes
(FWHM) comparable to the seeing size and thus have the brighter peaks
than those of the other objects having larger sizes.

Figure 3 shows the positions of the objects identified to be at
$z\sim5$ (filled circles) in the two color diagram 
($V - I_C$ and $I_C - z'$) as well as the unidentified objects with
$I_C<25.0$ mag (open triangles).
One of the seven LBGs shown in Figure 1 is undetected in 
$V$ band ($V>28.5$ mag), 
while the remaining six are detected in $V$ band within an aperture of
$1.^{\prime\prime}6$. 
Four among seven targets ($I_C<25.0$ mag) outside of the selection window  
are identified to be Galactic M stars (filled pentagons).
These results suggest that our selection criteria for objects at $z\sim5$ 
is reasonable.

\section{Discussion}

\citet{Shap03} divided spectroscopically confirmed $\sim$800 LBGs 
at $z\sim3$  into four subgroups (each group consists 
of $\sim$200 LBGs) based on  rest-frame Ly$\alpha$ equivalent width,
and made composite spectra for each group.
They found that about three fourth of the $\sim$ 800 LBGs at $z\sim3$ show 
Ly$\alpha$ emission lines,
and about two third of them show strong Ly$\alpha$ emission
(EW$_{\rm rest} > 10$\AA)
\footnote{\citet{Shap03} measured the equivalent width of Ly$\alpha$ line by summing
up both emission and absorption.
In our case, we measured the equivalent width of Ly$\alpha$ only for the 
emission part mainly because 
it is hard to distinguish intrinsic absorption and intergalactic
absorption. 
Thus the value of rest-frame equivalent width of Ly$\alpha$ by
\citet{Shap03} is the lower limit when it is compared with our equivalent width.}.
They also found that LBGs with weaker Ly$\alpha$ emission have stronger
LIS absorption lines, redder UV continuum slopes, and larger $E(B-V)$ values.
The weak Ly$\alpha$ emission, strong LIS absorption lines, the red UV color,
and the large $E(B-V)$ are considered to originate in dusty environment. 
In the LBGs with no or weak Ly$\alpha$ emission,
star formation may occur earlier and may be chemically more
evolved than those with strong Ly$\alpha$ emission.
Although Shapley et al. (2001) pointed out a
possibility that  the LBGs with weak Ly$\alpha$  emission 
are younger than those with strong emission from 
the SED fitting, this might be reconciled with 
the chemically evolved nature if the ages derived 
by SED fitting are affected by the most recent 
star formation occurred in the LBGs.

The spectra of our seven LBGs (expect for one AGN) at $z\sim5$ 
and thus the composite spectrum of them is significantly different
from the composite spectrum of LBGs at $z\sim3$ as shown in Figure 2.
The Ly$\alpha$ emission is much weaker, and the continuum depression in
the wavelength region shorter than the redshifted Ly$\alpha$ emission is
much larger than those at $z\sim3$.
The measured equivalent width of Ly$\alpha$ emission is $4.5$\AA\ at
$z\sim5$ while $15.1$\AA\ at $z\sim3$.
The LIS absorption lines are stronger in the spectrum for $z\sim5$ 
than in that for $z\sim3$; measured equivalent widths of \ion{Si}{2} 
$\lambda$1260, \ion{O}{1}+\ion{Si}{2} $\lambda$1303, and \ion{C}{2}
$\lambda$1334 are $-2.8$\AA, $-2.3$\AA, and $-2.4$\AA,
respectively for $z\sim5$, while $-1.7$\AA, $-2.3$\AA, and
$-1.5$\AA, respectively for $z\sim3$. 
However, the spectra of the seven LBGs at $z\sim5$ fairly
resemble to subpopulations of LBGs at $z\sim3$;
the composite spectra of LBGs at $z\sim3$ with
no or weak Ly$\alpha$ emission \citep[Group 1 and Group 2 by][]{Shap03}
are quite similar to the spectra of our LBGs at $z\sim5$.
The average rest-frame equivalent widths of
the three LIS absorption lines 
of the seven LBGs at $z\sim5$ is 
$-2.8$\AA\, being very close to $-2.5$\AA\ for Group 1 (no Ly$\alpha$
emission).
The average value of EW$_{\rm rest}$(LIS) corresponds to metallicity of
12$+$log(O/H)$\sim8.0$, if we assume that the relation obtained
in the local universe by \citet{Hek98} can be applied to the high
redshift LBGs, which is not certain at this moment.
These results suggest that these LBGs at $z\sim5$ are chemically
evolved to some degree.

All our LBGs confirmed to be at $z\sim5$ show no or
weak Ly$\alpha$ emission with relatively strong LIS absorption lines;
a fraction of LBGs with strong Ly$\alpha$ emission is very small,
though the sample size is still small. 
We may be witnessing some sign of evolution of LBGs from $z\sim5$ to
$z\sim3$. 
The lack of strong Ly$\alpha$ emission as well as the presence of strong
LIS absorption at $z\sim5$ are likely to be due to their dusty and
chemically evolved environment
(though the escape of Ly$\alpha$ photons may not be related in a simple way
to the metallicity of the galaxy \citep[e.g.,][]{kunth})
and to the presence of more neutral
hydrogen in and/or around a galaxy than that at $z\sim3$.

However, it is worth emphasizing that the LBGs we observed are
relatively brighter ones among those at $z\sim5$
($I_C = 25.0$ mag corresponds to $M^{\ast}$ of UVLF at $z\sim5$
\citep{Iwa03}.).
There is a possibility that the strength of Ly$\alpha$
emission depends on the magnitude (i.e., UV continuum).
\citet{Shap03} found for LBGs at $z\sim3$ that the average UV magnitude
is fainter for the LBGs with stronger Ly$\alpha$ emission.
They also found that the average rest-frame equivalent width of the Ly$\alpha$
emission line of faint LBGs is larger than that of bright LBGs
among the subgroup with strong Ly$\alpha$ emission of EW$_{\rm rest}
\geq$ 20\AA.
This trend could also be the case at $z\sim5$.
In fact, \citet{Lehn02}, who made a similar search for LBGs at $z\sim5$
using $R$, $I$, and $z$ band deep images,
found that all of their spectroscopically confirmed six objects to be at
$z\sim5$
show very strong 
(EW$_{\rm rest}>30$\AA) Ly$\alpha$ emission lines.
The $I$ magnitudes of them are
$\sim1$ magnitude fainter than those of our seven LBGs at $z\sim5$.
In addition, \citet{Ouchi03a} found that the number density of Ly$\alpha$
emitters (LAEs) at $z=4.86$ against to LBGs at $z\sim5$ rapidly
decreases with increasing
UV continuum light ($i'\lesssim25$ mag).
These observational results suggest that the brighter LBGs
tend to show the weaker Ly$\alpha$ emission also at $z\sim5$.
It is also worth noting here that the LBGs with secure redshifts tend to
show rather compact morphology.
Thus the weakness of Ly$\alpha$ emission and the strong LIS absorption 
lines may also relate to their morphological property, which
may also link to an evolutionary stage of galaxies.

If we interpret that the weakness of the Ly$\alpha$ emission
is caused by the dusty environment in the LBGs,
it might be possible that the brightest LBGs are the most chemically evolved
ones at the epoch.
It is known that brighter LBGs have a larger correlation length
at $z\sim3$ \citep{gd01}, suggesting that they are associated with more
massive dark halos and star formation may occur with biased manner in
the earlier epoch as compared with less clustered fainter LBGs.
The similar result has been obtained for LBGs at $z\sim4$ by
\citet{Ouchi03c}.
We also found the same clustering segregation with magnitude in our
sample of LBGs at $z\sim5$ (in preparation).
Thus 
the brighter LBGs at $z\sim5$ may be associated with even more massive
dark halos and have experienced more biased star formation, resulting in more
dusty environment as compared with many of LBGs at $z\sim3$
with a strong Ly$\alpha$ emission.

Another possible reason of the difference of Ly$\alpha$ appearance is the
effect of a strong clustering of LAEs.
\citet{Ouchi03a} and \citet{Shima} found that there is an overdensity
region ($\sim10^{\prime}\times10^{\prime}$) of LAEs (EW$_{\rm rest} > 14$\AA)
at $z\sim4.86$ from their wide-field deep narrow-band observations. 
The observed field (44 arcmin$^2$) of \citet{Lehn02} may happen to
fall on an overdensity region of Ly $\alpha$ emitters; their rest-frame
equivalent widths of Ly$\alpha$ emission ($>30-50$\AA)
is larger than the detection threshold of the LAE selection.
While 
our field of view for the spectroscopy  (total of $\sim$85 arcmin$^2$) may 
happen to point to a low density region of LAEs and we could not
observe LBGs with strong Ly$\alpha$ emission. 
However, since we selected spectroscopic targets from regions 
where the surface density of LBG candidates is relatively high,
this may be unlikely provided that the distribution of LBGs 
broadly coincides with that of LAEs at the same epoch.

To summarize,  the results presented here may show some sign of
evolution in spectroscopic feature from $z\sim5$ to $\sim3$,
or the presence of luminosity dependence of nature.
However, our sample size is too small to reach any significant conclusions.
Further spectroscopic observations of LBGs at $z\sim5$ over wider field
and magnitude range are necessary to reveal spectroscopic nature 
and discuss relationship with evolution of LBG population. 

\acknowledgments

We are grateful to the FOCAS team, especially support astronomer
Dr. Youichi Ohyama, for their advises and supports for our
observations. We thank all staffs of Subaru telescope. 
We also appreciate the referee for the comments which improved this paper.
II is supported by a Research Fellowship of the Japan Society for the
Promotion of Science for Young Scientists and 
by a Grant-in-Aid for the 21st Century Center of Excellence of Japan
"Center for Diversity and Universality in Physics".
This work is partly supported
by grants-in-aid for scientific research (15540233) from the Ministry of
Education, Culture, Sports, Science and Technology of Japan.

\begin{figure}
\figurenum{1}
\plotone{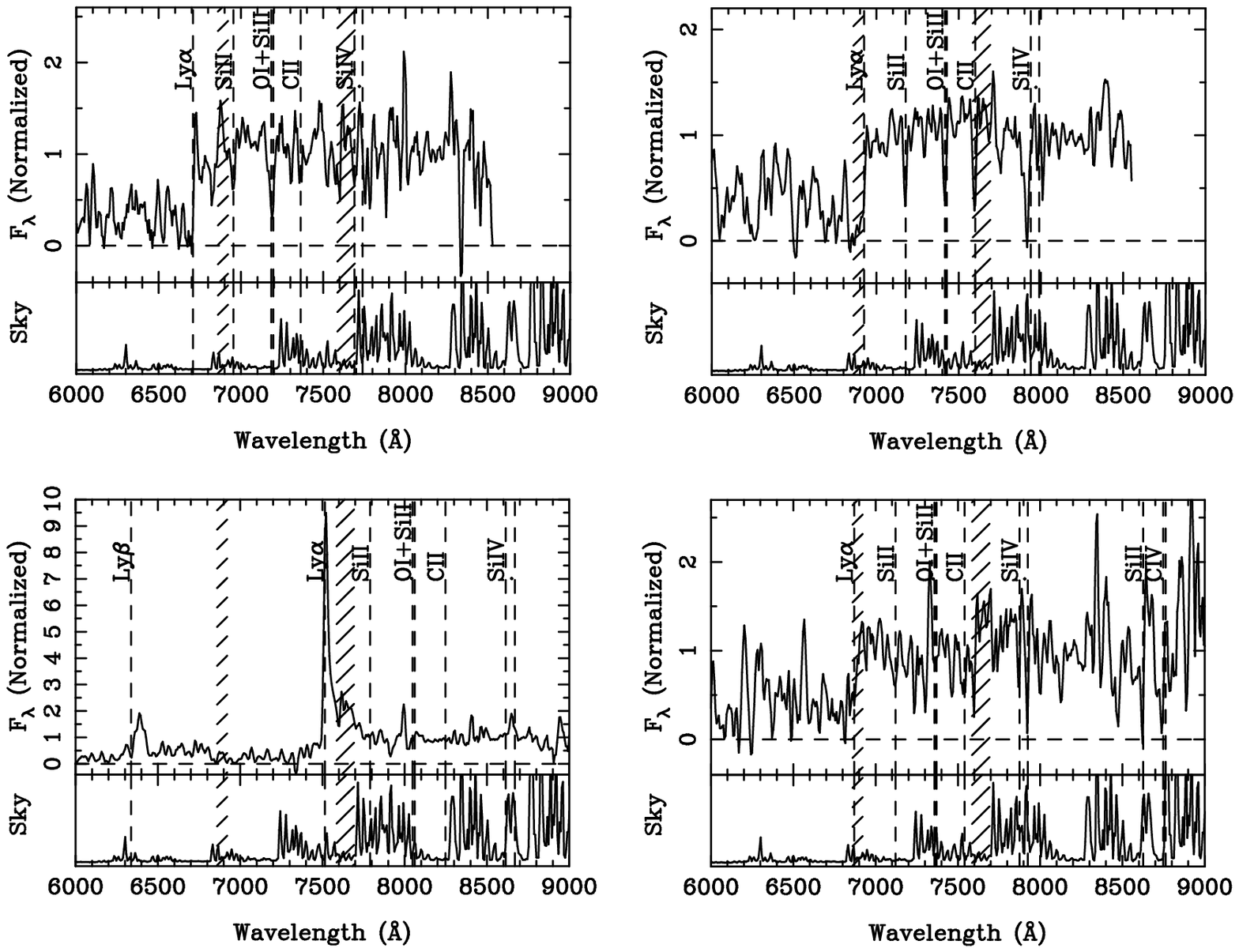}
\figcaption{Spectra of objects at $z\sim5$ with the positions of their rest-frame UV lines.
{\it Top}: Object No.1 ({\it left}) and No.2 ({\it right}). {\it Bottom}: Object No.3 ({\it left}: AGN of \citet{Bar02}) and No.4 ({\it right}). Flux scale is $F_{\lambda}$ and normalized with continuum level averaged over the region longer than Ly$\alpha$ excluding sky emission and LBGs absorption lines. These spectra are smoothed with the boxcar over 3-pixel. Sky spectrum is shown in a lower panel of each figure, and atmospheric absorptions are shown as vertical hatched regions.}
\end{figure}

\begin{figure}
\figurenum{1}
\plotone{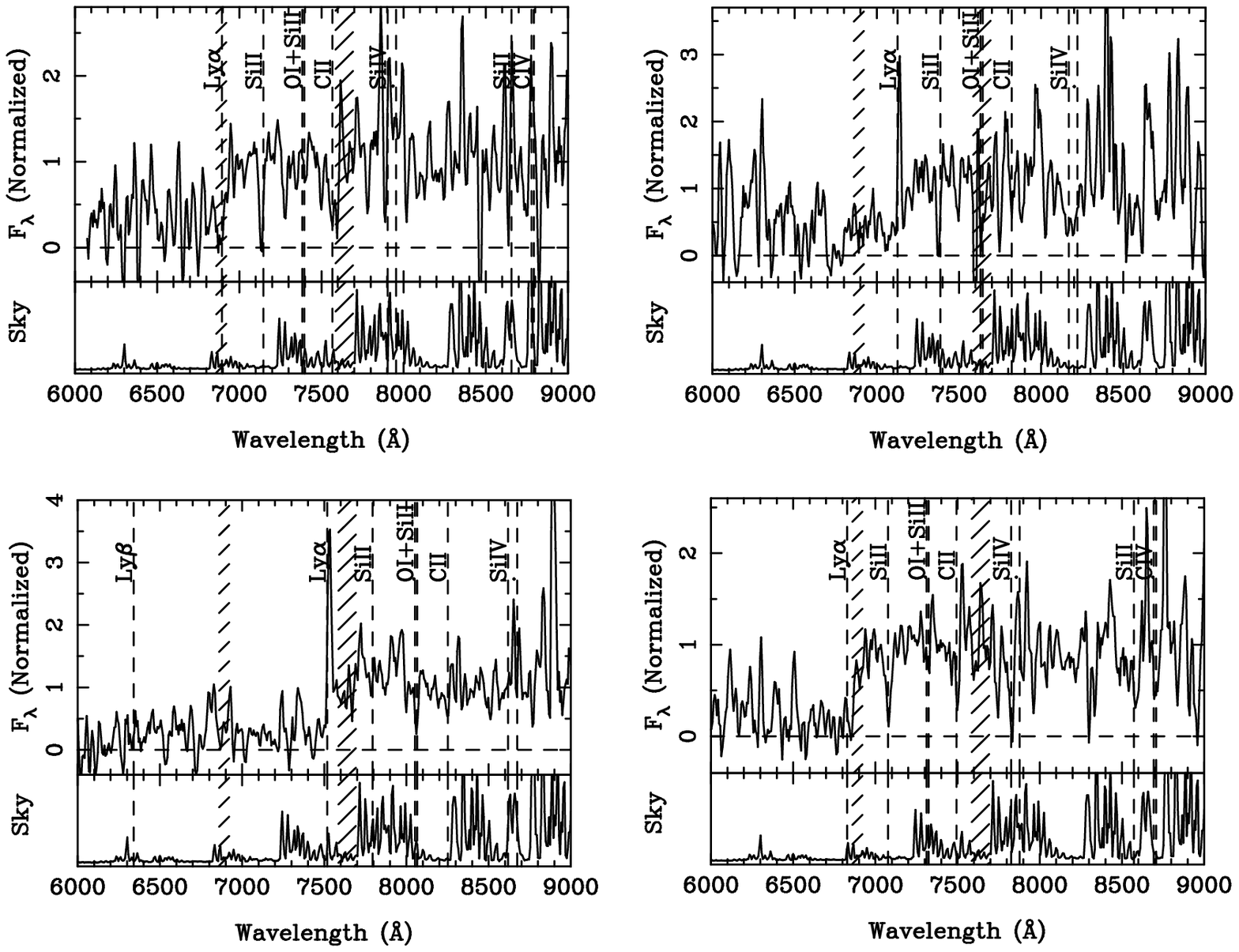}
\figcaption{Continued. {\it Top}: Object No.5 ({\it left}) and No.6 ({\it right}). {\it Bottom}: Object No.7 ({\it left}) and No.8 ({\it right}).}
\end{figure}

\begin{figure}
\figurenum{2}
\begin{center}
\plotone{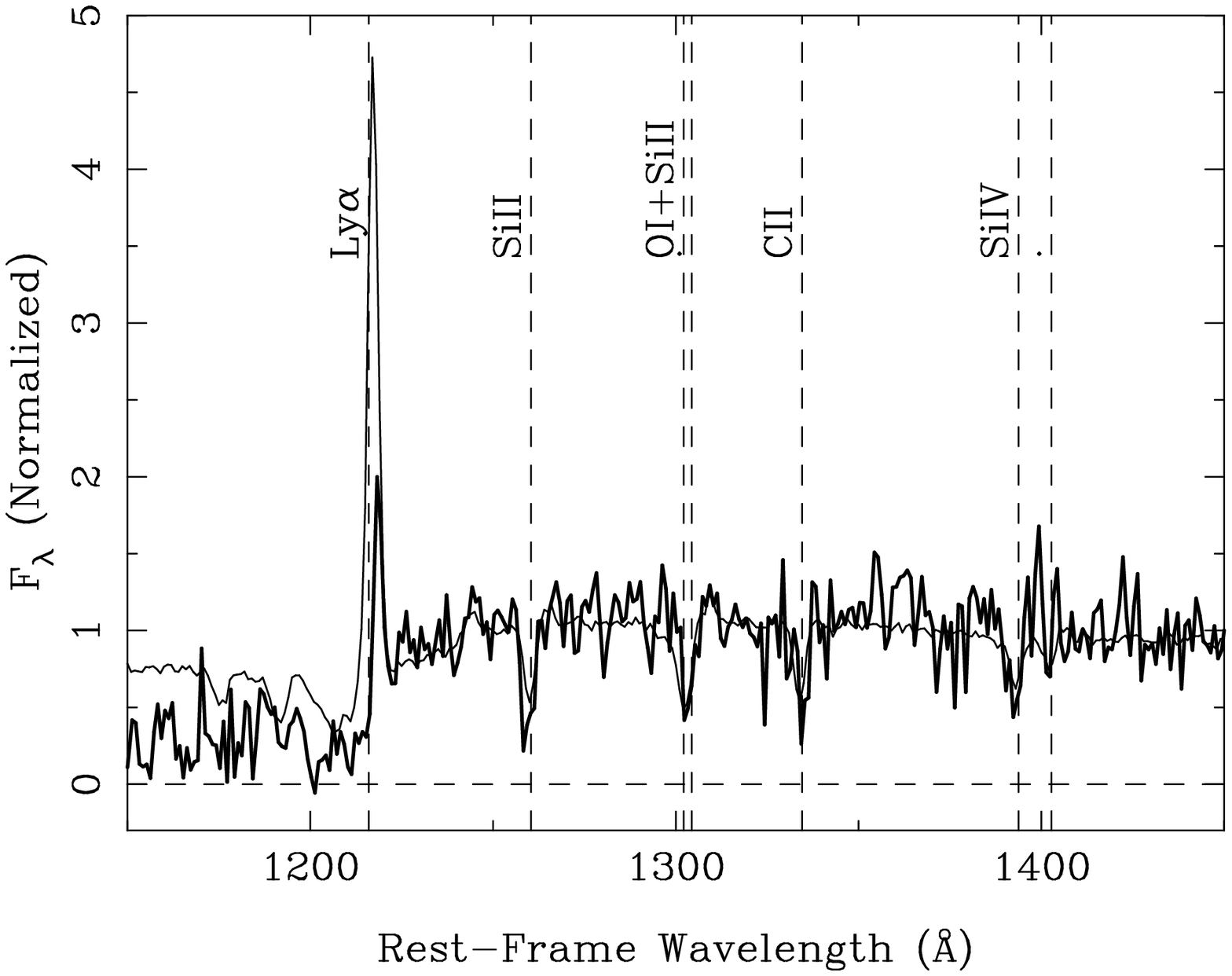}
\figcaption{Composite spectrum of the seven LBGs at $z\sim5$ (thick line). The composite spectrum of 811 LBGs at $z\sim3$ by \citet{Shap03} is over-plotted (thin line). Both of them are binned to a resolution of 1\AA\ per pixel.}
\end{center}
\end{figure}

\begin{center}
\begin{figure}
\figurenum{3}
\plotone{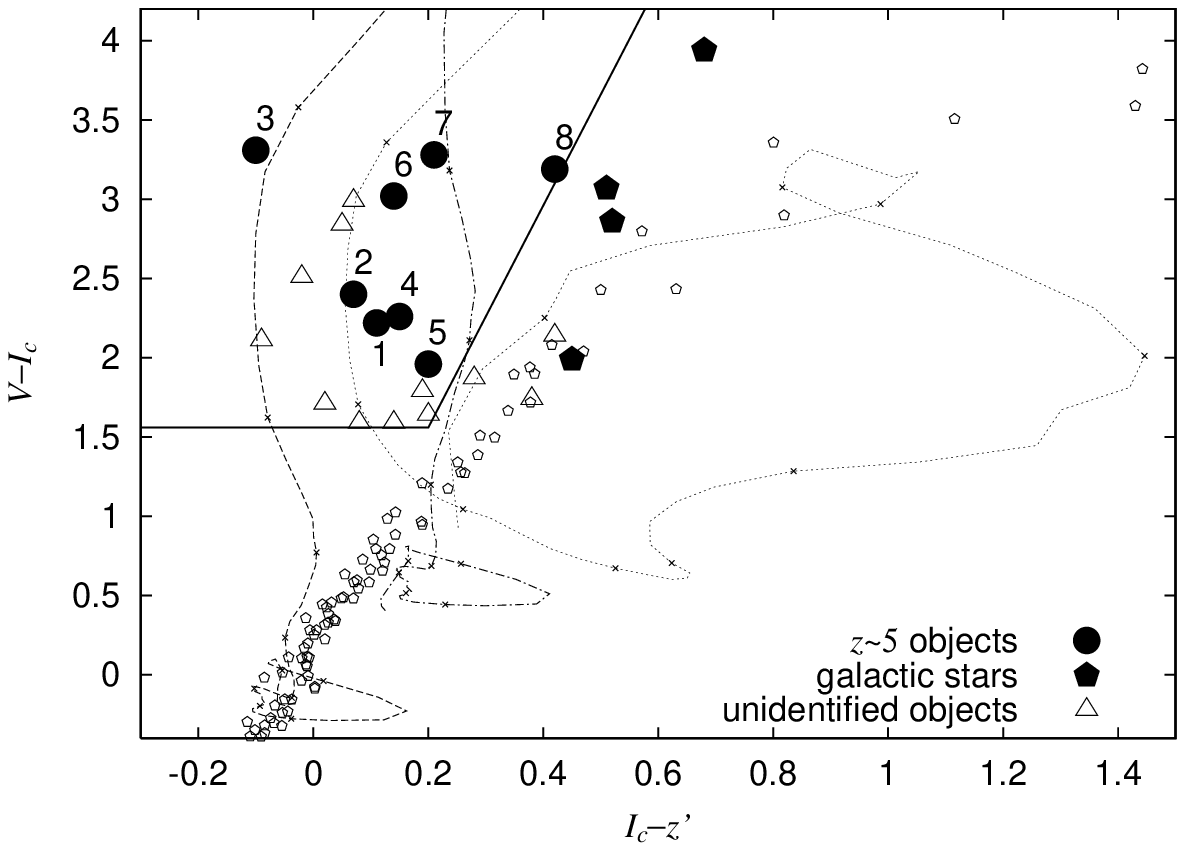}
\figcaption{Positions of our spectroscopic targets ($I_C<25.0$ mag) in two-color diagram. Our color selection criteria for LBGs at $z\sim5$, $V - I_C \geq 1.56$ mag and $\ V - I_C \geq 7 (I_C - z') + 0.16$ mag \citep{Iwa03}, are indicated by thick lines. 
Filled circles represent the objects confirmed to be at $z\sim5$ with each ID number shown in Table 1. 
Filled pentagons show objects identified to be Galactic M stars, and open triangles show unidentified objects.
A dashed (a dot-dashed) line represents a color track of a model LBG spectrum with the $E(B-V)=0.0$ mag ($E(B-V)=0.4$ mag) from \citet{Iwa03}. A dotted line refers to a color track of an elliptical galaxy (Coleman, Wu, \& Weedman 1980). The small cross symbols are plotted with a redshift interval of 0.5 from $z=0$ to $z=5$. No evolution is considered in these models. Small open pentagons indicate the colors of A0 -- M9 stars calculated based on the library by \citet{Pickles(1998)}.}
\end{figure}
\end{center}

\begin{deluxetable}{ccccccccc}
\tabletypesize{\scriptsize}
\tablecaption{Spectroscopic sample identified to be at $z\sim5$.\label{tbl-1}}
\tablewidth{0pt}
\tablehead{
\colhead{No. (ID)} & \colhead{R.A.(J2000)} & \colhead{Dec.(J2000)} & \colhead{$I_C$\tablenotemark{a}} & \colhead{$V-I_C$\tablenotemark{b}} & \colhead{$I_C-z'$\tablenotemark{b}} & \colhead{Redshift\tablenotemark{c}}& \colhead{EW(Ly$\alpha$)\tablenotemark{d,f}} & \colhead{EW(LIS)\tablenotemark{e,f}}}
\startdata
1 (F8-2) & 12 38 11.2 & +62 09 19.3 & 24.03 & 2.22 & 0.11 & 4.517 & 1.4 & $-2.2$\\
2 (F2-2) & 12 37 57.5 & +62 17 19.2 & 24.16 & 2.40 & 0.07 & 4.695 & 0 & $-2.2$\\
3\tablenotemark{g} (F1-3) & 12 36 47.9 & +62 09 41.3 & 24.17 & 3.31 & -0.10 & 5.186 & 44.4 & $-0.7$\\
4 (F1-1) & 12 37 05.7 & +62 07 43.1 & 24.62 & 2.26 & 0.15 & 4.650 & 0 & $-1.2$\\
5 (F2-6) & 12 38 29.0 & +62 16 18.8 & 24.63 & 1.96 & 0.20 & 4.667 & 0 & $-5.1$\\
6 (F2-3) & 12 38 25.5 & +62 18 19.7 & 24.74 & 3.02 & 0.14 & 4.857 & 6.3 & $-2.7$\\
7 (F2-8) & 12 38 04.4 & +62 14 19.8 & 24.80 & 3.28 & 0.21 & 5.183 & 9.8 & $-2.0$\\
8 (F2-7) & 12 38 16.7 & +62 18 05.5 & 24.83 & $>$3.19 & 0.42 & 4.615 & 0 & $-3.9$\\
\enddata
\tablenotetext{a}{$I_C$ magnitude. MAG\_AUTO from SExtractor \citep{Bet} is adopted. Photometric errors are estimated to be 0.16 mag and 0.19 mag for $24.0$ mag$ < I_C \leq 24.5$ mag and $24.5$ mag$ < I_C \leq 25.0$ mag, respectively.}
\tablenotetext{b}{These values are measured with a $1^{\prime\prime}.6$ aperture. The errors in $V-I_C$ and $I_C - z'$ colors are $0.05-0.2$ mag and $0.05-0.1$ mag, respectively, depending on the $I_C$ magnitude and color.}
\tablenotetext{c}{Redshifts determined as an average of the two or three low-ionized interstellar metal absorption lines (Si II $\lambda$1260, O I+Si II $\lambda$1303, and C II $\lambda$1334) except for the object No.3; we adopted the redshift measured with Ly$\alpha$ emission for this object. The error of redshift is $\sim0.006$ (comparable to the spectral resolution).}
\tablenotetext{d}{Rest-frame equivalent width of Ly$\alpha$ emission. A value zero refers to an object without a Ly$\alpha$ emission.}
\tablenotetext{e}{Rest-frame equivalent width of the average of three LIS absorption lines (Si II $\lambda$1260, O I+Si II $\lambda$1303, and C II $\lambda$1334).}
\tablenotetext{f}{Error of equivalent width for each line is estimated to be $20\sim30$\%.} 
\tablenotetext{g}{X-ray selected AGN \citep{Bar02}.}
\tablecomments{Units of right ascension are hours, minutes, and seconds, and units of declination are degrees, arcminutes, and arcseconds.}
\end{deluxetable}

\end{document}